\documentclass[12pt]{article}
\usepackage{epsfig}
\usepackage{amsmath}
\usepackage{hhline}
\usepackage{amssymb}
\usepackage{times}
\usepackage{cite}
\usepackage{psfrag}
\usepackage{color}
\usepackage{here}
\usepackage{changebar}

\usepackage[pagewise]{lineno}

\newlength{\dinwidth}
\newlength{\dinmargin}
\begin{document}  
%\linenumbers
% The rest
\newcommand{\pom}{{I\!\!P}}
\newcommand{\reg}{{I\!\!R}}
\newcommand{\slowpi}{\pi_{\mathit{slow}}}
\newcommand{\fiidiii}{F_2^{D(3)}}
\newcommand{\fiidiiiarg}{\fiidiii\,(\beta,\,Q^2,\,x)}
\newcommand{\n}{1.19\pm 0.06 (stat.) \pm0.07 (syst.)}
\newcommand{\nz}{1.30\pm 0.08 (stat.)^{+0.08}_{-0.14} (syst.)}
\newcommand{\fiidiiiful}{F_2^{D(4)}\,(\beta,\,Q^2,\,x,\,t)}
\newcommand{\fiipom}{\tilde F_2^D}
\newcommand{\ALPHA}{1.10\pm0.03 (stat.) \pm0.04 (syst.)}
\newcommand{\ALPHAZ}{1.15\pm0.04 (stat.)^{+0.04}_{-0.07} (syst.)}
\newcommand{\fiipomarg}{\fiipom\,(\beta,\,Q^2)}
\newcommand{\pomflux}{f_{\pom / p}}
\newcommand{\nxpom}{1.19\pm 0.06 (stat.) \pm0.07 (syst.)}
\newcommand {\gapprox}
   {\raisebox{-0.7ex}{$\stackrel {\textstyle>}{\sim}$}}
\newcommand {\lapprox}
   {\raisebox{-0.7ex}{$\stackrel {\textstyle<}{\sim}$}}
\def\gsim{\,\lower.25ex\hbox{$\scriptstyle\sim$}\kern-1.30ex%
\raise 0.55ex\hbox{$\scriptstyle >$}\,}
\def\lsim{\,\lower.25ex\hbox{$\scriptstyle\sim$}\kern-1.30ex%
\raise 0.55ex\hbox{$\scriptstyle <$}\,}
\newcommand{\pomfluxarg}{f_{\pom / p}\,(x_\pom)}
\newcommand{\dsf}{\mbox{$F_2^{D(3)}$}}
\newcommand{\dsfva}{\mbox{$F_2^{D(3)}(\beta,Q^2,x_{I\!\!P})$}}
\newcommand{\dsfvb}{\mbox{$F_2^{D(3)}(\beta,Q^2,x)$}}
\newcommand{\dsfpom}{$F_2^{I\!\!P}$}
\newcommand{\gap}{\stackrel{>}{\sim}}
\newcommand{\lap}{\stackrel{<}{\sim}}
\newcommand{\fem}{$F_2^{em}$}
\newcommand{\tsnmp}{$\tilde{\sigma}_{NC}(e^{\mp})$}
\newcommand{\tsnm}{$\tilde{\sigma}_{NC}(e^-)$}
\newcommand{\tsnp}{$\tilde{\sigma}_{NC}(e^+)$}
\newcommand{\st}{$\star$}
\newcommand{\sst}{$\star \star$}
\newcommand{\ssst}{$\star \star \star$}
\newcommand{\sssst}{$\star \star \star \star$}
\newcommand{\tw}{\theta_W}
\newcommand{\sw}{\sin{\theta_W}}
\newcommand{\cw}{\cos{\theta_W}}
\newcommand{\sww}{\sin^2{\theta_W}}
\newcommand{\cww}{\cos^2{\theta_W}}
\newcommand{\trm}{m_{\perp}}
\newcommand{\trp}{p_{\perp}}
\newcommand{\trmm}{m_{\perp}^2}
\newcommand{\trpp}{p_{\perp}^2}
\newcommand{\alp}{\alpha_s}

\newcommand{\alps}{\alpha_s}
\newcommand{\sqrts}{$\sqrt{s}$}
\newcommand{\LO}{$O(\alpha_s^0)$}
\newcommand{\Oa}{$O(\alpha_s)$}
\newcommand{\Oaa}{$O(\alpha_s^2)$}
\newcommand{\PT}{p_{\perp}}
\newcommand{\JPSI}{J/\psi}
\newcommand{\sh}{\hat{s}}
\newcommand{\uh}{\hat{u}}
\newcommand{\MP}{m_{J/\psi}}
\newcommand{\PO}{I\!\!P}
\newcommand{\xbj}{x}
\newcommand{\xpom}{x_{\PO}}
\newcommand{\ttbs}{\char'134}
\newcommand{\xpomlo}{3\times10^{-4}}  
\newcommand{\xpomup}{0.05}  
\newcommand{\dgr}{^\circ}
\newcommand{\pbarnt}{\,\mbox{{\rm pb$^{-1}$}}}
\newcommand{\gev}{\,\mbox{GeV}}
\newcommand{\WBoson}{\mbox{$W$}}
\newcommand{\fbarn}{\,\mbox{{\rm fb}}}
\newcommand{\fbarnt}{\,\mbox{{\rm fb$^{-1}$}}}
%
% Some useful tex commands
%
\newcommand{\qsq}{\ensuremath{Q^2} }
\newcommand{\gevsq}{\ensuremath{\mathrm{GeV}^2} }
\newcommand{\et}{\ensuremath{E_t^*} }
\newcommand{\rap}{\ensuremath{\eta^*} }
\newcommand{\gp}{\ensuremath{\gamma^*}p }
\newcommand{\dsiget}{\ensuremath{{\rm d}\sigma_{ep}/{\rm d}E_t^*} }
\newcommand{\dsigrap}{\ensuremath{{\rm d}\sigma_{ep}/{\rm d}\eta^*} }

% Some more useful tex commands
%
\newcommand{\GeV}{\rm GeV}
\newcommand{\TeV}{\rm TeV}
\newcommand{\pb}{\rm pb}
\newcommand{\cm}{\rm cm}
\newcommand{\hdick}{\noalign{\hrule height1.4pt}}
\newcommand{\ptc}{\ensuremath{p_t^{\mathrm{calo}}}}
\newcommand{\ptm}{\ensuremath{p_t^{\mathrm{miss}}}}
\newcommand{\ptg}{\ensuremath{p_t^{\gamma}}}
\newcommand{\ptj}{\ensuremath{p_t^{\mathrm{jet}}}}
\newcommand{\rpv}{\mbox{$R_p \!\!\!\!\!\! / \;\; $}}
\newcommand{\grav}{\tilde{G}}
\newcommand{\sell}{\tilde{e}_L}
\newcommand{\neu}{\tilde{\chi}_1^0}
\newcommand{\epm}{$e^\pm$}
\newcommand{\ev}{\,\mbox{eV}}

% Journal macro
\def\Journal#1#2#3#4{{#1} {\bf #2} (#3) #4}
\def\NCA{\em Nuovo Cimento}
\def\NIM{\em Nucl. Instrum. Methods}
\def\NIMA{{\em Nucl. Instrum. Methods} {\bf A}}
\def\NPB{{\em Nucl. Phys.}   {\bf B}}
\def\PLB{{\em Phys. Lett.}   {\bf B}}
\def\PRL{\em Phys. Rev. Lett.}
\def\PRD{{\em Phys. Rev.}    {\bf D}}
\def\ZPC{{\em Z. Phys.}      {\bf C}}
\def\EJC{{\em Eur. Phys. J.} {\bf C}}
\def\CPC{\em Comp. Phys. Commun.}

\begin{titlepage}

\begin{flushleft}
DESY 04-227 \hfill ISSN 0418-9833 \\
November 2004
\end{flushleft}

\vspace{2cm}

\begin{center}
\begin{Large}

{\bf Search for Light Gravitinos in Events with Photons and Missing
Transverse Momentum at HERA}

\vspace{2cm}

H1 Collaboration

\end{Large}
\end{center}

\vspace{2cm}

%%%%%%%%%%%%%%%%%%%%%%%%%%%%%%%%%%%%%%%%%%%%%%%%%%%%%%%%%%%%%%%%%%%%%%%%
%%%%%%%%%%%%%%%%%%%%%%%%%%%%%%%%%%%%%%%%%%%%%%%%%%%%%%%%%%%%%%%%%%%%%%%%
\begin{abstract}
%%%%%%%%%%%%%%%%%%%%%%%%%%%%%%%%%%%%%%%%%%%%%%%%%%%%%%%%%%%%%%%%%%%%%%%%
%%%%%%%%%%%%%%%%%%%%%%%%%%%%%%%%%%%%%%%%%%%%%%%%%%%%%%%%%%%%%%%%%%%%%%%%

A search for gravitinos produced in \epm$p$ collisions is
performed using the 
H1 detector at HERA. 
The data were taken at a centre-of-mass energy of $319\,\GeV$ and
correspond to an integrated luminosity of $64.3\,\mathrm{pb^{-1}}$ for
$e^+p$ collisions and $13.5\,\mathrm{pb^{-1}}$ for $e^-p$ collisions.
If $R$-parity is not conserved, the $t$-channel exchange of a
selectron can produce a neutralino, which, in models where the gravitino is
the lightest supersymmetric particle, subsequently decays into a
photon and a light gravitino. The resulting event signature, which involves 
an isolated photon, a jet and missing transverse            
energy, is analysed for the first time at HERA.
No deviation from the Standard Model is found. 
Exclusion limits on the cross section and on $R$-parity-violating Yukawa
couplings are
derived in a Gauge Mediated Supersymmetry Breaking scenario. The results
are independent of the squark 
sector. 
Neutralinos and supersymmetric partners of the left-handed electron with
masses up to $112\,\GeV$ and $164\,\GeV$, 
respectively, can be ruled out
at the $95\,\%$ confidence level for $R$-parity-violating couplings
$\lambda'$ equal to 1.

\end{abstract}

\vspace{1.5cm}

\begin{center}
To be submitted to {{\em Phys. Lett.}   {\bf B}}
\end{center}

\end{titlepage}

%
%          COPY THE AUTHOR AND INSTITUTE LISTS 
%       AT THE TIME OF THE T0-TALK INTO YOUR AREA
%
% from /h1/iww/ipublications/h1auts.tex 

\begin{flushleft}
  %-- H1AUTS Author list by names 
%-- Status: Tue Feb  3 18:26:54 MET 2004  Number of authors = 310 

A.~Aktas$^{10}$,               %DESY-ST        03/2            Aktas               
V.~Andreev$^{26}$,             %LPI -PD        8/88            Andreev             
T.~Anthonis$^{4}$,             %ANTW-ST        11/99           Anthonis            
A.~Asmone$^{33}$,              %ROME-ST        07/2            Asmone              
A.~Babaev$^{25}$,              %ITEP-PD        8/88            Babaev              
S.~Backovic$^{37}$,            %ZEUT-PD        03/2            Backovic            
J.~B\"ahr$^{37}$,              %ZEUT-PD        8/88            Baehr               
P.~Baranov$^{26}$,             %LPI -PD        8/88            Baranovp            
E.~Barrelet$^{30}$,            %PARI-PD        11/99           Barrelet            
W.~Bartel$^{10}$
% ...

%-- H1 Institutes 
\bigskip{\it
% ...
$ ^{4}$ Inter-University Institute for High Energies ULB-VUB, Brussels;
         Universiteit Antwerpen, Antwerpen; Belgium$^{ c}$ \\
% ...
$ ^{10}$ DESY, Hamburg, Germany \\
% ...
$ ^{25}$ Institute for Theoretical and Experimental Physics,
          Moscow, Russia$^{ l}$ \\
 $ ^{26}$ Lebedev Physical Institute, Moscow, Russia$^{ e}$ \\
% ...
$ ^{30}$ LPNHE, Universit\'{e}s Paris VI and VII, IN2P3-CNRS,
          Paris, France \\
% ...
$ ^{33}$ Dipartimento di Fisica Universit\`a di Roma Tre
          and INFN Roma~3, Roma, Italy \\
% ...
$ ^{37}$ DESY, Zeuthen, Germany \\
% ...
$ ^c$ Supported by FNRS-FWO-Vlaanderen, IISN-IIKW and IWT
     and  by Interuniversity Attraction Poles Programme,
     Belgian Science Policy \\
% ...
$ ^e$ Supported by the Deutsche Forschungsgemeinschaft \\
% ...
$ ^l$ Partially Supported by Russian Foundation
      for Basic Research, grant    no. 00-15-96584 \\
}

\end{flushleft}

\newpage

%%%%%%%%%%%%%%%%%%%%%%%%%%%%%%%%%%%%%%%%%%%%%%%%%%%%%%%%%%%%%%%%%%%%%%%%
%%%%%%%%%%%%%%%%%%%%%%%%%%%%%%%%%%%%%%%%%%%%%%%%%%%%%%%%%%%%%%%%%%%%%%%%
\section{Introduction}
%%%%%%%%%%%%%%%%%%%%%%%%%%%%%%%%%%%%%%%%%%%%%%%%%%%%%%%%%%%%%%%%%%%%%%%%
%%%%%%%%%%%%%%%%%%%%%%%%%%%%%%%%%%%%%%%%%%%%%%%%%%%%%%%%%%%%%%%%%%%%%%%%

Supersymmetry (SUSY) \cite{susy} is an attractive concept which remedies
some shortcomings 
of the Standard Model (SM). This fermion-boson symmetry leads to an extension of 
the particle spectrum by associating to each SM particle a supersymmetric 
partner, differing in its spin by half a unit. The masses of the new particles
are related to the symmetry breaking mechanism. 
In Gauge Mediated Supersymmetry Breaking (GMSB) models, new
``messenger'' fields are introduced which couple to the source of
supersymmetry breaking. The breaking is then transmitted to the SM fields
and their superpartners by gauge interactions~\cite{gmsbrev}. 
The gravitino, $\tilde{G}$, is the lightest
supersymmetric particle
(LSP) and can be as light as $10^{-3}\,\ev$.

\begin{figure}[hhh]
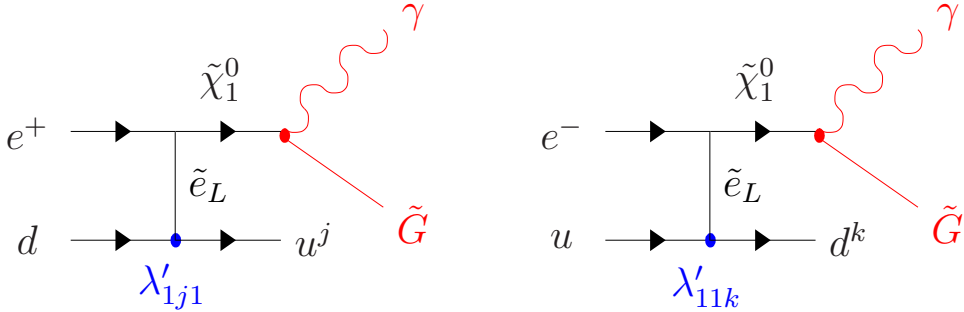
 
  \begin{center}
   \psfrag{a}[][][1.3][0]{$e^+$}
   \psfrag{b}[][][1.3][0]{$d$}
   \psfrag{c}[][][1.3][0]{$\neu$}
   \psfrag{d}[][][1.3][0]{$u^j$}
   \psfrag{k}[][][1.3][0]{\color{blue} $\lambda'_{1j1}$}
   \psfrag{s}[][][1.3][0]{\color{blue} $c$}
   \psfrag{t}[][][1.3][0]{\color{black} $\tilde{e}_L$}
   \psfrag{gr}[][][1.3][0]{\color{red} $\tilde{G}$}
   \psfrag{ga}[][][1.3][0]{\color{red} $\gamma$}
   \epsfig{file=fig1.ps,width=7cm}
   \psfrag{a}[][][1.3][0]{$e^-$}
   \psfrag{b}[][][1.3][0]{$u$}
   \psfrag{c}[][][1.3][0]{$\neu$}
   \psfrag{d}[][][1.3][0]{$d^k$}
   \psfrag{k}[][][1.3][0]{\color{blue} $\lambda'_{11k}$}
   \psfrag{s}[][][1.3][0]{\color{blue} $c$}
   \psfrag{t}[][][1.3][0]{\color{black} $\tilde{e}_L$}
   \psfrag{gr}[][][1.3][0]{\color{red} $\tilde{G}$}
   \psfrag{ga}[][][1.3][0]{\color{red} $\gamma$}
   \epsfig{file=fig1.ps,width=7cm}
%
%\put(-160,166){\sf (a)}
%\put(42,193){\sf (b)}
%
%
  \end{center}
  \caption{Dominant diagrams for neutralino production via \rpv\ selectron
   exchange in $e^+p$ and $e^-p$ scattering, with subsequent neutralino decay into
    a gravitino and a photon.}
  \label{fig:feyn}
\end{figure}

The next-to-lightest supersymmetric particle (NLSP) is generally
either the lightest neutralino $\neu$ or a slepton
$\tilde{\ell}$, which decays to the stable gravitino via $\neu \rightarrow
\gamma \tilde{G}$ or 
$\tilde{\ell} \rightarrow \ell \tilde{G}$. 
The distinguishing event topology involves a
high energy photon or lepton and significant missing energy due to the
undetected gravitino. Such topologies have been studied
at LEP \cite{lepgmsb} and the Tevatron
\cite{Abbott:1997rd,Abazov:2004jx}. No significant deviation from the SM was
found. In these studies $R$-parity ($R_p$) was assumed to be
conserved. 
An investigation of $R_p$-violating (\rpv) SUSY in
a GMSB scenario is performed in this analysis.
A search for \rpv\ resonant
single neutralino production $\neu$ via $t$-channel selectron exchange, $e^\pm q
\rightarrow \neu q'$, is performed in $e^+p$ and $e^-p$ collisions. 
It is assumed that the $\neu$ is  
the NLSP and that the decay $\neu
\rightarrow \gamma \tilde{G}$ occurs with an unobservably
small lifetime and dominates over \rpv\
neutralino decays.
Feynman diagrams of the analysed processes are depicted in
Fig.~\ref{fig:feyn}. The resulting
experimental signature is a photon, a jet originating from the
scattered quark and 
missing transverse momentum due to the escaping gravitino. The main SM
background arises from radiative charged current (CC) deep inelastic
scattering (DIS) with a jet, a photon and a neutrino in the final state. 
%
%%%%%%%%%%%%%%%%%%begin%%figure%%%%%%%%%%%%%%%%%%%%%%%%%%%%%%%%%%%%%%%
%%%%%%%%%%%%%%%%%%%%%%%%%%%%%%end%%figure%%%%%%%%%%%%%%%%%%%%%%%%%%%%%

Resonant squark production in \rpv\ 
SUSY
has been investigated previously at HERA
in models in which the LSP is either a gaugino \cite{rpvhera} or a light
squark \cite{Aktas:2004tm}. Squark mass dependent limits 
on various \rpv\ Yukawa couplings have been derived. In contrast, the
process considered in this analysis is completely independent of the squark
sector. 

The data correspond to an integrated luminosity of $64.3\,\pb^{-1}$ for
$e^+p$ collisions recorded in 1999 and 2000 and $13.5\,\pb^{-1}$ for
$e^-p$ collisions recorded in 1998 and 1999. The energy of the incoming
electron\footnote{In the 
  following {\em 
    electron} will be used to refer to both electron and positron unless
  explicitly otherwise stated.} is $E_e=27.6\,\GeV$ and the energy of the
incoming proton is
$E_p=920\,\GeV$. Thus the electron-proton centre-of-mass energy is
$319\,\GeV$. 

%%%%%%%%%%%%%%%%%%%%%%%%%%%%%%%%%%%%%%%%%%%%%%%%%%%%%%%%%%%%%%%%%%%%%%%%
%%%%%%%%%%%%%%%%%%%%%%%%%%%%%%%%%%%%%%%%%%%%%%%%%%%%%%%%%%%%%%%%%%%%%%%%
\section{The Supersymmetric Model}
%%%%%%%%%%%%%%%%%%%%%%%%%%%%%%%%%%%%%%%%%%%%%%%%%%%%%%%%%%%%%%%%%%%%%%%%
%%%%%%%%%%%%%%%%%%%%%%%%%%%%%%%%%%%%%%%%%%%%%%%%%%%%%%%%%%%%%%%%%%%%%%%%

%%
This analysis considers a supersymmetric model where the gravitino is the
LSP and in which $R_p$ is not conserved -- a
scenario which has been used e.g. in \cite{Allanach:2001hx} and has been
considered before in the context of dark matter \cite{darkmatter}.
$R$-parity is a discrete multiplicative symmetry which can be written as
$R_p=(-1)^{3B+L+2S}$, where $B$ denotes the baryon number, $L$ the lepton
number and $S$ the spin of a particle. 
The most general supersymmetric theory that is renormalisable and gauge
invariant with respect to the SM gauge group \cite{wein}
contains \rpv\ Yukawa 
couplings between the supersymmetric partner of the left-handed electron
$\tilde{e}_L$, a
left-handed 
up-type quark $u_L^j$ and a right-handed down-type anti-quark
$\bar{d}_R^k$, where $j$ and $k$ denote generation indices. The
corresponding part of the Lagrangian reads
\begin{equation}
  {\cal L}_{\rpv} = - \lambda'_{1jk} \tilde{e}_L u_L^j \bar{d}_R^k + {\rm h.c.}
  \label{eq:rpv}
\end{equation}
At HERA, the presence of couplings $\lambda'_{1j1}$ and $\lambda'_{11k}$
could lead to neutralino production 
in $e^+p$ and $e^-p$ collisions, respectively, via $t$-channel selectron
exchange (see 
Fig.~\ref{fig:feyn}). The search presented here is performed under the
simplifying 
assumption that one of the couplings $\lambda'_{1j1}$ ($j=1,2$)
or
$\lambda'_{11k}$ ($k=1,2,3$)
dominates\footnote{The
coupling $\lambda'_{131}$ is not studied here because the production of
a top quark together with a neutralino is suppressed due to the high top
quark mass.}.  
If the initial state lepton is a positron the
dominant hard scattering process at the large Bjorken $x$ values relevant
here involves a
down quark from the 
proton (see Fig.~\ref{fig:feyn}, left). If
the initial state lepton is an electron mainly up quarks are
probed (see Fig.~\ref{fig:feyn}, right). 
For a given \rpv\ coupling, the $\neu$ production cross section
for an initial electron is roughly a factor of two larger than that
for an initial positron, reflecting the different parton
densities for valence up and down quarks in the proton.
Due to the contribution of diagrams involving antiquarks in the
initial state (not shown in Fig.~\ref{fig:feyn}), the cross section for $\neu$
production in $e^+p$ ($e^-p$) collisions via a $\lambda'_{111}$ coupling is
larger than that for production via a $\lambda'_{121}$ ($\lambda'_{112}$) coupling
of the same strength. The relative difference amounts to at most
$15\%$ ($8\%$) for $e^+p$ ($e^-p$) processes, for low masses of the produced
neutralino. The cross sections of the $e^-p$ processes induced by
$\lambda'_{112}$ and $\lambda'_{113}$ are the same within a few percent.

The GMSB model used here is inspired by
\cite{Dimopoulos:1996yq}. While the gaugino mass spectrum and gauge
couplings are derived from this minimal model, the 
slepton masses are treated as free parameters.
The supersymmetric partner of the left-handed electron
can be much lighter than 
the supersymmetric partner of the right-handed one
as,
for example, in the 
Hybrid Multi-Scale Supersymmetric Model HMSSM-I \cite{Ambrosanio:1997ky}.
This allows small mass differences $\Delta m =
m({\tilde{e}_L})-m({\neu})$.

The GMSB model is characterised by
six new parameters
in addition to those of the SM:
\begin{equation}
  \Lambda,\; M,\; N,\; \tan\beta,\; {\rm sign}(\mu)\; {\text{and}}\; \sqrt{F} \,.
\end{equation}
The parameter $\Lambda$ sets the overall mass scale for the SUSY particles, $M$
is the mass of the messenger particles,
$N$ is the number of
sets of messenger particles, 
$\tan\beta$ is the ratio of the Higgs vacuum expectation values and
${\rm sign}(\mu)$ is the sign of the Higgs sector mixing parameter
$\mu$.
The intrinsic SUSY breaking scale is $\sqrt{F}$, which also determines the 
$\grav$ mass according to $m_{\tilde{G}} \simeq 2.5 \cdot F/(100\ {\rm TeV})^2
\, {\rm eV}$. 
Furthermore, $\sqrt{F}$ affects the neutralino decay rate according to
$\Gamma(\neu \rightarrow \gamma \tilde{G}) \sim 1/F^2$; low values of
$\sqrt{F}$ thus suppress the \rpv\ decays of the neutralino. In the SUSY
parameter space considered here, the branching ratio
${\rm BR}(\neu \rightarrow 
\gamma \tilde{G})$ exceeds $95 \,\%$  if $\sqrt{F}$ lies in the range
between the 
present experimental limit of $221\,\GeV$ \cite{Acosta:2002eq} and
$1\,\TeV$. Thus, $\sqrt{F}$ is not varied but it is assumed that the 
\rpv\ decays of the neutralino do not contribute. 
In the range considered for $\sqrt{F}$ the neutralino lifetime is short
enough to 
have no effect on the detection efficiency. 
The contributions of the heavier neutralinos 
$\tilde{\chi}^0_i$ ($i=2,3,4$) to the considered signal
are small and thus neglected.

%%%%%%%%%%%%%%%%%%%%%%%%%%%%%%%%%%%%%%%%%%%%%%%%%%%%%%%%%%%%%%%%%%%%%%%%
%%%%%%%%%%%%%%%%%%%%%%%%%%%%%%%%%%%%%%%%%%%%%%%%%%%%%%%%%%%%%%%%%%%%%%%%
\section{The H1 Detector}
%%%%%%%%%%%%%%%%%%%%%%%%%%%%%%%%%%%%%%%%%%%%%%%%%%%%%%%%%%%%%%%%%%%%%%%%
%%%%%%%%%%%%%%%%%%%%%%%%%%%%%%%%%%%%%%%%%%%%%%%%%%%%%%%%%%%%%%%%%%%%%%%%

In the following the detector components most
relevant for this analysis are briefly described. 
The main components of the tracking system are the central drift and
proportional chambers which cover the polar angle\footnote{The polar angle
  $\theta$ is measured with respect to the proton beam direction.}
range $20^{\circ} < \theta < 160^{\circ}$ and a forward track detector
($7^{\circ} < \theta < 
25^{\circ}$).
The tracking system is
surrounded by a finely segmented liquid argon (LAr)
calorimeter~\cite{Andrieu:1993kh} 
which covers the range
$4^{\circ} < \theta <154^{\circ}$ and which has an
energy resolution of
 $\sigma_E / E \simeq 12\,\% / \sqrt{E({\rm GeV})} \oplus 1\,\%$ for
 electrons and $ \sigma_E / E \simeq 50\,\% / \sqrt{E({\rm GeV})} \oplus 2\,\%$ for
 hadrons, as obtained in test beam measurements~\cite{Andrieu:1994yn}. The
 tracking system and calorimeters are surrounded by a 
 superconducting solenoid and its iron yoke instrumented with streamer tubes.
The latter are used to detect hadronic showers which extend beyond the LAr
and to identify muons. 
The luminosity is
determined from the rate of Bethe-Heitler events ($ep \rightarrow
ep\gamma$) measured in a luminosity monitor. 
A detailed description of the H1 experiment can be found
in~\cite{Abt:1997hi}. 

%%%%%%%%%%%%%%%%%%%%%%%%%%%%%%%%%%%%%%%%%%%%%%%%%%%%%%%%%%%%%%%%%%%%%%%%
%%%%%%%%%%%%%%%%%%%%%%%%%%%%%%%%%%%%%%%%%%%%%%%%%%%%%%%%%%%%%%%%%%%%%%%%
\section{Event Simulation}
%%%%%%%%%%%%%%%%%%%%%%%%%%%%%%%%%%%%%%%%%%%%%%%%%%%%%%%%%%%%%%%%%%%%%%%%
%%%%%%%%%%%%%%%%%%%%%%%%%%%%%%%%%%%%%%%%%%%%%%%%%%%%%%%%%%%%%%%%%%%%%%%%

In order to estimate the expected SM contributions to the signature under
study and to determine the
detection efficiencies for a possible SUSY signal, complete Monte Carlo (MC)
simulations of the H1 detector response are performed.
For each possible SM source a sample of MC events is used
corresponding to a luminosity of more than 10 times 
that of the data.
For the simulation of the charged and neutral current (CC and NC) DIS
backgrounds,  
the DJANGO \cite{DJANGO} event generator is used which includes first order
QED radiation as modelled by HERACLES \cite{HERACLES}.
The parton densities in the proton are taken
from the CTEQ5L~\cite{CTEQ} parameterisation. 
The direct and resolved photoproduction
of light and heavy quark flavours
is generated using the PYTHIA~\cite{PYTHIA} program.
The SM predictions for $ep \rightarrow e Z X$ and $ep \rightarrow e W^\pm X$ 
are calculated using the leading order generator EPVEC~\cite{EPVEC} with
the next-to-leading order QCD corrections implemented using a reweighting
method \cite{wnlo}.

% SIGNAL SIMULATION:

The signal topology is simulated using the SUSYGEN generator~\cite{SUSYGEN}.
The parton densities are evaluated
at the scale of the Mandelstam variable $-t$. Efficiencies are determined by
interpolation between calculations at different points in the parameter
space, where the neutralino mass $m({\neu})$ is varied from
$50\,\GeV$ to $140\,\GeV$ and the selectron mass $m(\sell)$
from $m({\neu})$ to $200\,\GeV$, both in steps
of typically $15\,\GeV$.

%%%%%%%%%%%%%%%%%%%%%%%%%%%%%%%%%%%%%%%%%%%%%%%%%%%%%%%%%%%%%%%%%%%%%%%%
%%%%%%%%%%%%%%%%%%%%%%%%%%%%%%%%%%%%%%%%%%%%%%%%%%%%%%%%%%%%%%%%%%%%%%%%
\section{Search for the Process \boldmath{$ e^\pm q \rightarrow \neu q'
    \rightarrow \gamma \tilde{G} q' $}}
\label{seq:selection}
%%%%%%%%%%%%%%%%%%%%%%%%%%%%%%%%%%%%%%%%%%%%%%%%%%%%%%%%%%%%%%%%%%%%%%%%
%%%%%%%%%%%%%%%%%%%%%%%%%%%%%%%%%%%%%%%%%%%%%%%%%%%%%%%%%%%%%%%%%%%%%%%%

%%%%%%%%%%%%%%%%%%%%%%%%%%%%%%%%%%%%%%%%%%%%%%%%%%%%%%%%%%%%%%%%%%%%%%%%
\subsection{Event Preselection}
\label{seq:preselection}
%%%%%%%%%%%%%%%%%%%%%%%%%%%%%%%%%%%%%%%%%%%%%%%%%%%%%%%%%%%%%%%%%%%%%%%%

The process $ e^\pm q \rightarrow \neu q' \rightarrow \gamma \tilde{G} q'$ is
characterised 
by missing transverse energy,  a jet and an electromagnetic cluster in the
calorimeter. 
The events used in this analysis are triggered by the
LAr system with an efficiency of typically
$95\,\%$ for the chosen kinematic region.
Background events not related to $ep$ collisions are suppressed by
requiring a primary interaction vertex reconstructed within
$\pm35\,\mbox{cm}$ in $z$ of the nominal vertex position, by using
topological filters against cosmic and proton-beam related background and
by requiring an event time which is consistent with the bunch crossing
time. 

Events are selected if the missing transverse momentum determined from the
energy 
deposits in the calorimeter
is greater than 
$25\,\GeV$. 
The events are required to contain at least one hadronic jet
in the range $10^{\circ} < \theta_{\rm jet} < 145^{\circ}$ 
and an identified photon with a polar angle $\theta_\gamma$ greater than
$10^{\circ}$, both with transverse momenta greater than $5\,\GeV$.
Hadronic 
jets are reconstructed from energy deposits in the calorimeter
using a cone algorithm in the laboratory frame with a radius $\sqrt{
  (\Delta \eta)^2 + (\Delta \phi)^2 } = 1$, where $\eta = -\ln \tan
{\theta}/{2}$ is the pseudorapidity and $\phi$ denotes the azimuthal angle.

Photons 
are identified using a shower shape analysis of energy deposits in
the LAr calorimeter.
For $\theta_\gamma > 20^{\circ}$ an electromagnetic cluster
is only accepted 
as a photon candidate
if it is not associated with a charged track in the central tracking
system.
In addition, the photon must not lie within the cone of any
reconstructed jet with $p_{T,{\rm jet}} > 5\,\GeV$.

%%%%%%%%%%%%%%%%%%%%%%%%%%%%%%%%%%%%%%%%%%%%%%%%%%%%%%%%%%%%%%%%%%%%%%%%
\subsection{Systematic Uncertainties}
%%%%%%%%%%%%%%%%%%%%%%%%%%%%%%%%%%%%%%%%%%%%%%%%%%%%%%%%%%%%%%%%%%%%%%%%
%
The systematic errors on the SM background expectation are evaluated by
considering the following uncertainties.
\begin{itemize}
\item The uncertainty on the electromagnetic energy scale of the
calorimeter varies from $0.7\,\%$ to $3\,\%$ depending on the calorimeter
region~\cite{Adloff:2003uh}. 
\item For the jet transverse momenta selected in this analysis (typically
  above $ 20\,\GeV$) the uncertainty on the hadronic energy
  scale is $2\,\%$~\cite{Adloff:2002au}.
 \item The uncertainty on the track reconstruction efficiency is
   $2\,\%$.
 \item An uncertainty of $10 \,\%$ is attributed to the SM cross sections
   for CC and NC DIS as implemented in the MC simulation which arises
   mainly from the parton densities of the proton at high $x$.
 \item The measurement of the integrated luminosity has a precision of
 $1.5\,\%$. 
\end{itemize}
Furthermore, the following uncertainties related to the modelling of
the SUSY signal are 
taken into account.
\begin{itemize}
 \item The theoretical uncertainty of the signal cross section due to 
   the uncertainty of the parton densities, which is typically a few percent
   and does not exceed $7\,\%$ for $e^-p$ scattering or
   $17\,\%$ for $e^+p$ scattering anywhere in the parameter space studied.
 \item Choosing either the invariant mass of the final state particles or
   the transverse momentum of the final state quark instead of the square
   root of 
   the Mandelstam variable $-t$ as the hard scale at which the parton
   distributions are evaluated yields an additional
   theoretical uncertainty of up to $10\,\%$ at large selectron and neutralino
   masses.
 \item A relative uncertainty of $ 10 \,\%$ is attributed to the signal
   detection 
   efficiencies, resulting mainly from the interpolation between the
   neutralino and selectron masses.
\end{itemize}

All systematic errors are added in quadrature separately for the signal and
the background. 
The resulting uncertainties are between $11\,\%$ and
$22\,\%$ in both cases.

%%%%%%%%%%%%%%%%%%%%%%%%%%%%%%%%%%%%%%%%%%%%%%%%%%%%%%%%%%%%%%%%%%%%%%%%
\subsection{Final Selection and Results}
\begin{figure}[hhh] 
\vspace{-11cm}
  \begin{center}
\begin{picture}(0,100)
   \put(-85,-10){\epsfig{file=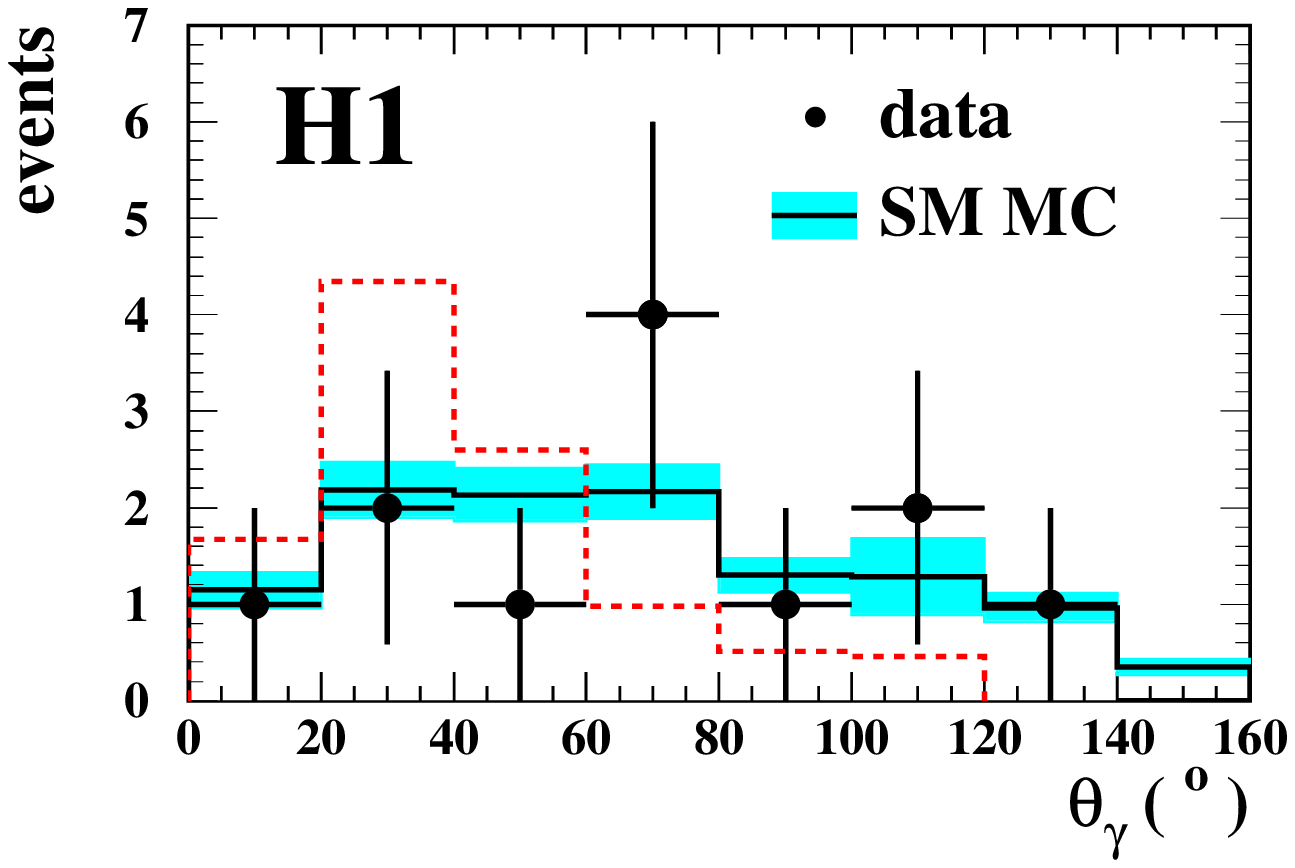,width=8cm,bbllx=129,bblly=660,bburx=501,bbury=389}}
   \put(0,-10){\epsfig{file=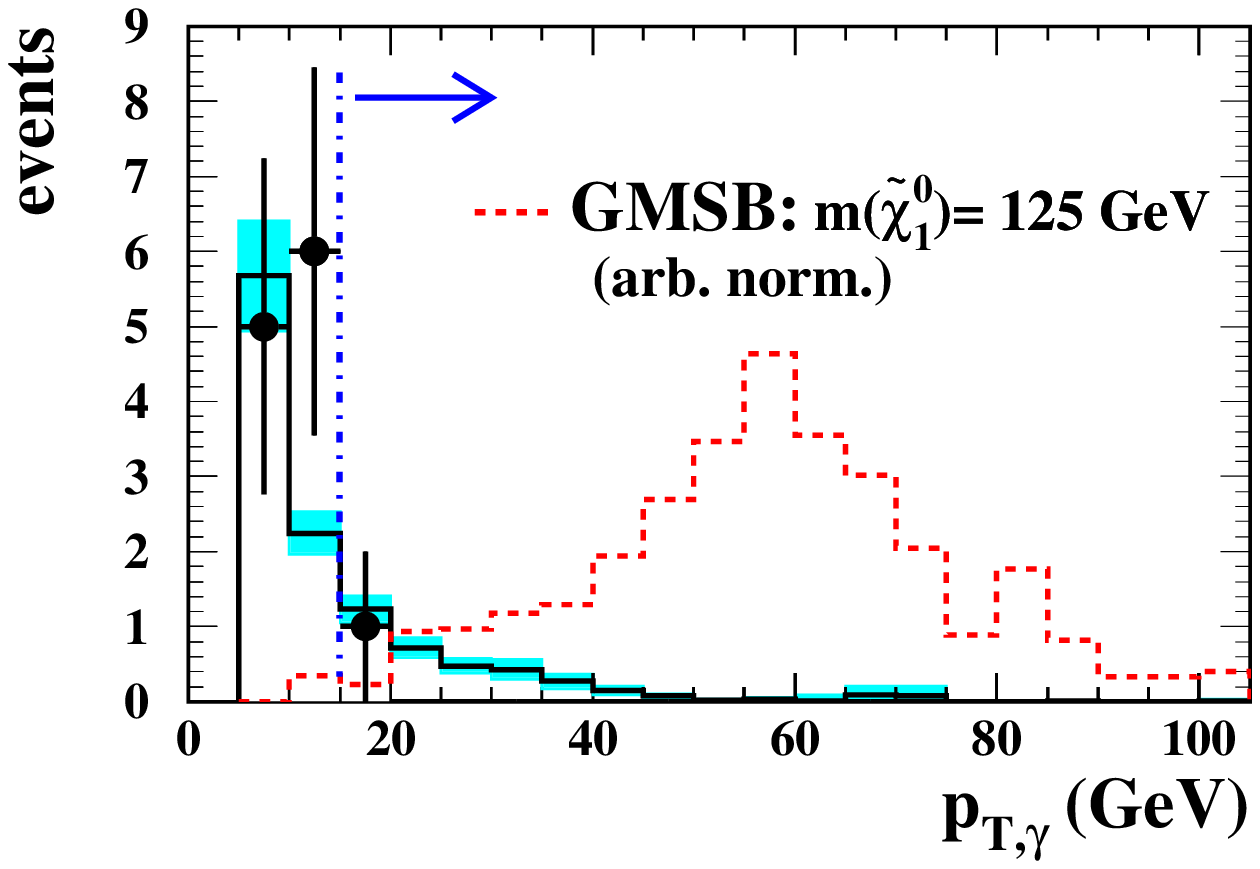,width=8cm,bbllx=129,bblly=660,bburx=501,bbury=389}}
   \put(-85,-70){\epsfig{file=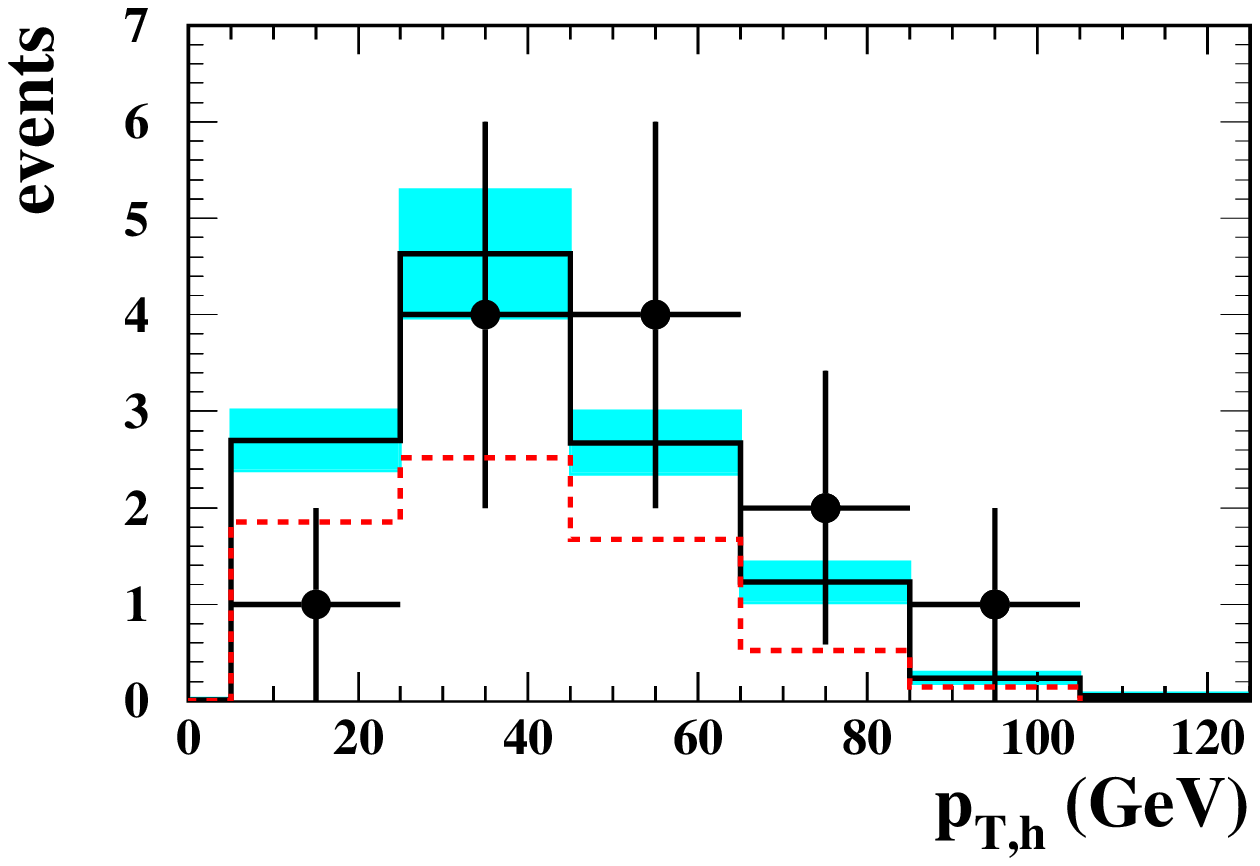,width=8cm,bbllx=129,bblly=660,bburx=501,bbury=389}}
   \put(0,-70){\epsfig{file=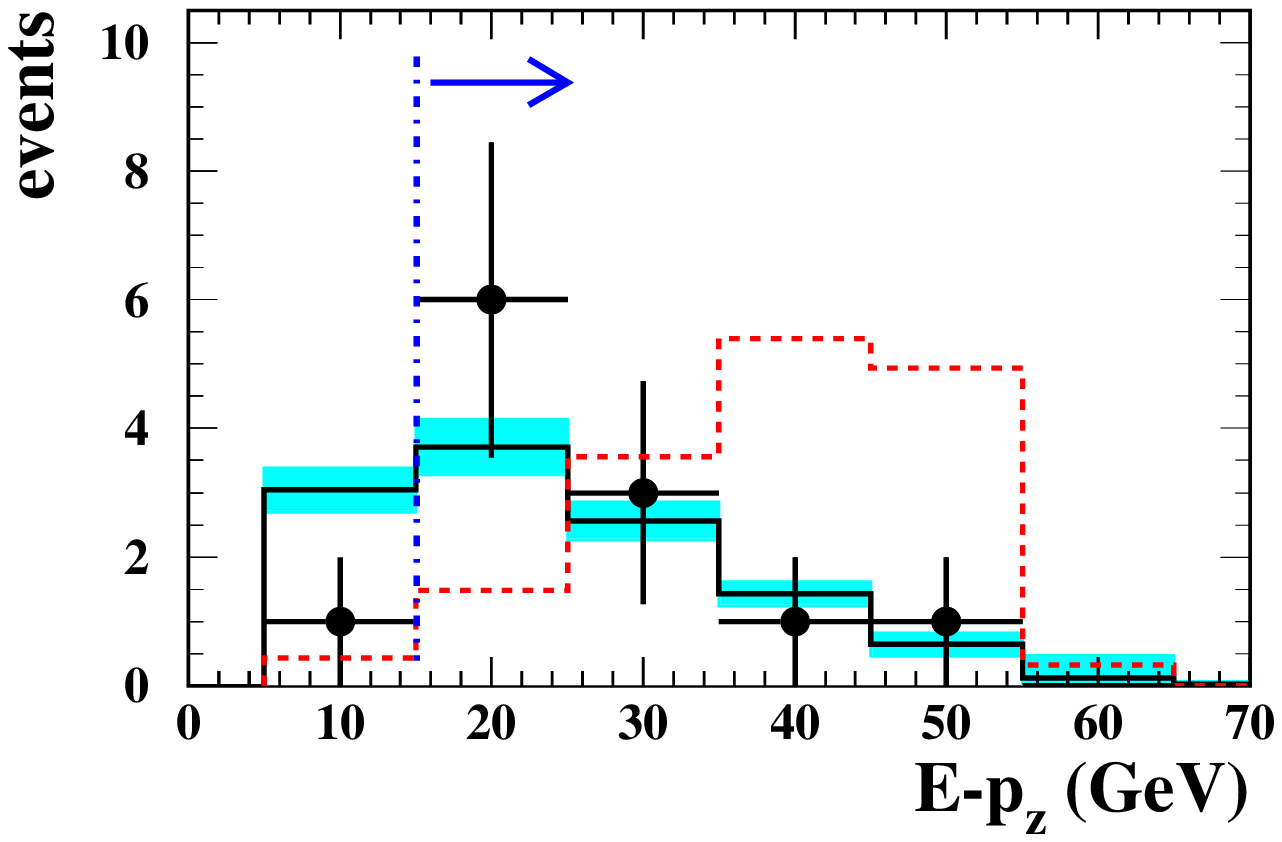,width=8cm,bbllx=129,bblly=660,bburx=501,bbury=389}}
\put(-15.5,-20){\sf (a)}
\put(69.5,-20){\sf (b)}
\put(-15.5,-80){\sf (c)}
\put(69.5,-80){\sf (d)}
\end{picture}
  \end{center}
  \vspace{12cm}
  \caption{Distributions of the polar angle (a) and transverse momentum (b) of photon
    candidates, hadronic 
    transverse momentum (c)
    and the sum of the $E-p_{z}$ of all measured particles (d) 
    after preselection. The complete
    $e^{\pm}p$ data set is compared with the SM
    prediction. The signal expected for a neutralino with a mass
    of $125\,\GeV$ is
    shown 
    with arbitrary normalisation (dashed histogram). The arrows indicate
    additional cuts
    applied on $p_{T,\gamma}$ and $E-p_z$ in
    the final selection.}
  \label{fig:ptg}
\end{figure}

%%%%%%%%%%%%%%%%%%%%%%%%%%%%%%%%%%%%%%%%%%%%%%%%%%%%%%%%%%%%%%%%%%%%%%%%
%
After the preselection, described in Section~\ref{seq:preselection}, 12
candidate events are selected  in the
complete $e^{\pm} p$ 
data sample 
and $11.5\pm 1.5$ events are expected from SM background processes,
predominantly from radiative CC DIS ($95\,\%$). 
The distributions of the polar angle $\theta_\gamma$ and the
transverse momentum $p_{T,\gamma}$ of the photon candidates are 
shown in Fig.~\ref{fig:ptg} (a) and (b), respectively. 
Fig.~\ref{fig:ptg} (c) shows the transverse momentum $p_{T,h}$
calculated from the hadronic energy deposits in the calorimeter. 
The sum of the 
$E-p_{z}$ 
of all measured particles is presented in Fig.~\ref{fig:ptg}
(d). 
The distributions illustrate
the good understanding of the SM processes. For comparison, a simulated SUSY
signal for a $\neu$ mass of $125\,\GeV$ is also shown.

To reduce
the CC DIS background,
$p_{T,\gamma} > 15\,\GeV$ and 
$E-p_z > 15\,\GeV$ are required for the final selection. These cuts are
also depicted in
Fig.~\ref{fig:ptg}.
%
%%%%%%%%%%%%%%%%%%begin%%figure%%%%%%%%%%%%%%%%%%%%%%%%%%%%%%%%%%%%%%%
%%%%%%%%%%%%%%%%%%%%%%%%%%%%%%end%%figure%%%%%%%%%%%%%%%%%%%%%%%%%%%%%
%
No candidate event is found in the $e^+p$ data set,
to be compared with $1.8\pm0.2$ expected from SM processes.
In the $e^-p$ data sample, 1 candidate event is found while the SM
prediction is  
$ 1.1 \pm 0.2 $.
The SM expectation arises predominantly from CC~DIS ($90\,\%$) with small
contributions 
from NC~DIS and the production of $W$ and $Z$ bosons where the final state
electron is misidentified as a photon.
With all cuts applied, the final selection efficiency for the signal
ranges between $10\,\%$ for low and $35\,\%$ for high
neutralino masses. 
The largest contribution
to the inefficiency arises from the
missing transverse energy requirement.   

\begin{figure}[hhh] 
\vspace{-11cm}
  \begin{center}
\begin{picture}(0,100)
   \put(-58,-10){\epsfig{file=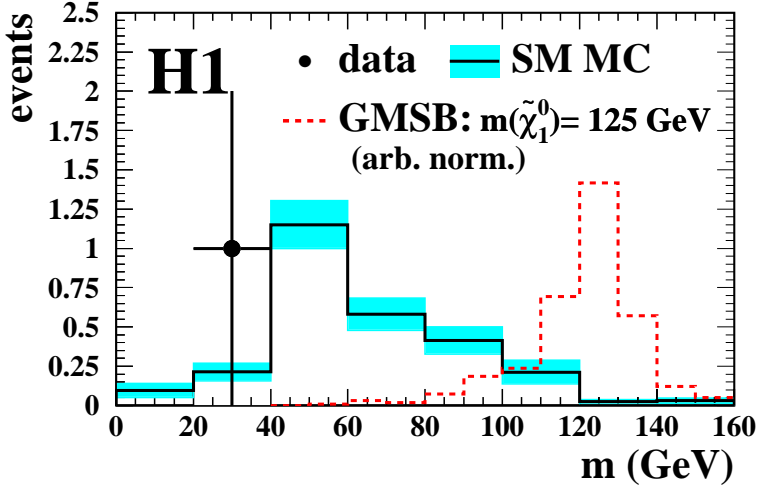,width=10cm,bbllx=129,bblly=660,bburx=501,bbury=389}}
%
%\Large
%\put(34.5,-26){$\bf\color{red}\rightarrow$}
%\put(42,193){\sf (b)}
%
\end{picture}
  \end{center}
  \vspace{7.3cm}
  \caption{Distribution of the invariant mass of the 
    photon candidate and the reconstructed missing particle in the final
    selection. The complete 
    $e^{\pm}p$ data set is compared with the SM prediction. The signal
    expected for a neutralino with a mass 
    of $125\,\GeV$ is
    shown 
    with arbitrary normalisation (dashed histogram).}
  \label{fig:mchi0}
\end{figure}

Assuming that the massless gravitino is the only non-interacting particle
in the event, its kinematics are reconstructed by exploiting the conservation
of transverse momentum and the constraint $(E-p_z) + (E_{\tilde{G}} - p_{z, \tilde{G}}) = 2E_e$.
The four-vector of this particle is then added to that of the 
photon to reconstruct the invariant mass $m$ of the decaying neutralino.
The data and the SM expectation
for this distribution are shown in Fig.~\ref{fig:mchi0}.
From the simulation of the SUSY signal, also shown in Fig.~\ref{fig:mchi0},
the mass resolution is determined to be around $10\,\GeV$.
The candidate event has a reconstructed invariant neutralino mass
of $36\pm 4\,\GeV$. 
%
%%%%%%%%%%%%%%%%%%begin%%figure%%%%%%%%%%%%%%%%%%%%%%%%%%%%%%%%%%%%%%%
%%%%%%%%%%%%%%%%%%%%%%%%%%%%%%end%%figure%%%%%%%%%%%%%%%%%%%%%%%%%%%%%

%%%%%%%%%%%%%%%%%%%%%%%%%%%%%%%%%%%%%%%%%%%%%%%%%%%%%%%%%%%%%%%%%%%%%%%%
%%%%%%%%%%%%%%%%%%%%%%%%%%%%%%%%%%%%%%%%%%%%%%%%%%%%%%%%%%%%%%%%%%%%%%%%
\section{GMSB Model Interpretations}
%%%%%%%%%%%%%%%%%%%%%%%%%%%%%%%%%%%%%%%%%%%%%%%%%%%%%%%%%%%%%%%%%%%%%%%%
%%%%%%%%%%%%%%%%%%%%%%%%%%%%%%%%%%%%%%%%%%%%%%%%%%%%%%%%%%%%%%%%%%%%%%%%

As no significant deviation from the SM is observed, constraints on GMSB 
models at the $95\,\%$ confidence level (CL) are derived using a
modified 
frequentist approach based on likelihood ratios, which takes statistical
and systematic uncertainties into account~\cite{Junk:1999kv}.
For a given neutralino mass $m(\neu)$, the limits are obtained by counting
the number of observed and expected events in a certain mass
interval. 
In the investigated range of $50\,{\rm GeV} < m(\neu) < 140\,\GeV$, a
mass interval of 
$\pm 2$ standard deviations, 
varying linearly between $\pm 16.5\,\GeV$ and $\pm 30\,\GeV$ around
$m(\neu)$, is chosen. 

For the interpretation of the results, the GMSB parameters $\tan\beta$, $N$
and ${\rm sign}(\mu)$
are fixed. The neutralino mass is scanned by varying 
$\Lambda$ at fixed $M/\Lambda$, the masses being
calculated using
the SUSPECT program~\cite{SUSPECT}. The 
mass of the supersymmetric partner of the left-handed electron 
is taken as a free
parameter. All other sfermions are assumed to be heavy.
Two example scenarios are considered. In the
first, the masses considered correspond to a scan of the parameters in the
range $30\,\TeV \le \Lambda \le 100\,\TeV$ taking $M/\Lambda = 2$  
for $\tan\beta=2$ and $N=1$ with negative
$\mu$.
In the second scenario, $\tan\beta=6$, $N=2$, $\mu<0$ and the parameter
range is $20\,\TeV \le
\Lambda \le 50\,\TeV$ with $M/\Lambda = 10^3$.
For a given 
neutralino mass, a variation of $N$ has only a minor effect on the
cross section, whereas the cross section
decreases significantly with increasing $\tan\beta$.

\begin{figure}[hhh] 
\vspace{-11cm}
  \begin{center}
\begin{picture}(0,100)
         \put(-82,-15){\epsfig{file=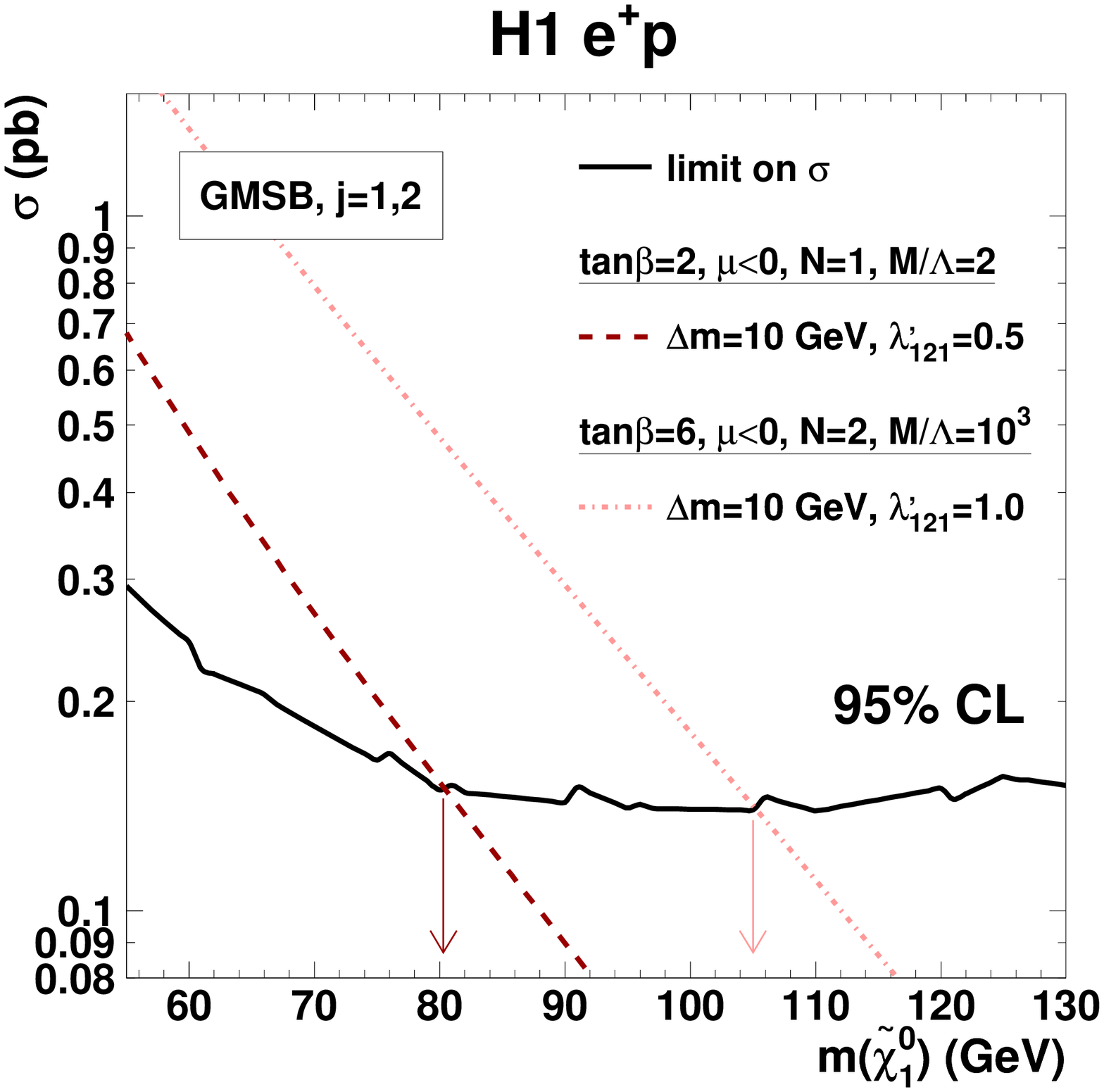,width=8cm,bbllx=25,bblly=660,bburx=575,bbury=130}}
         \put(3,-15){\epsfig{file=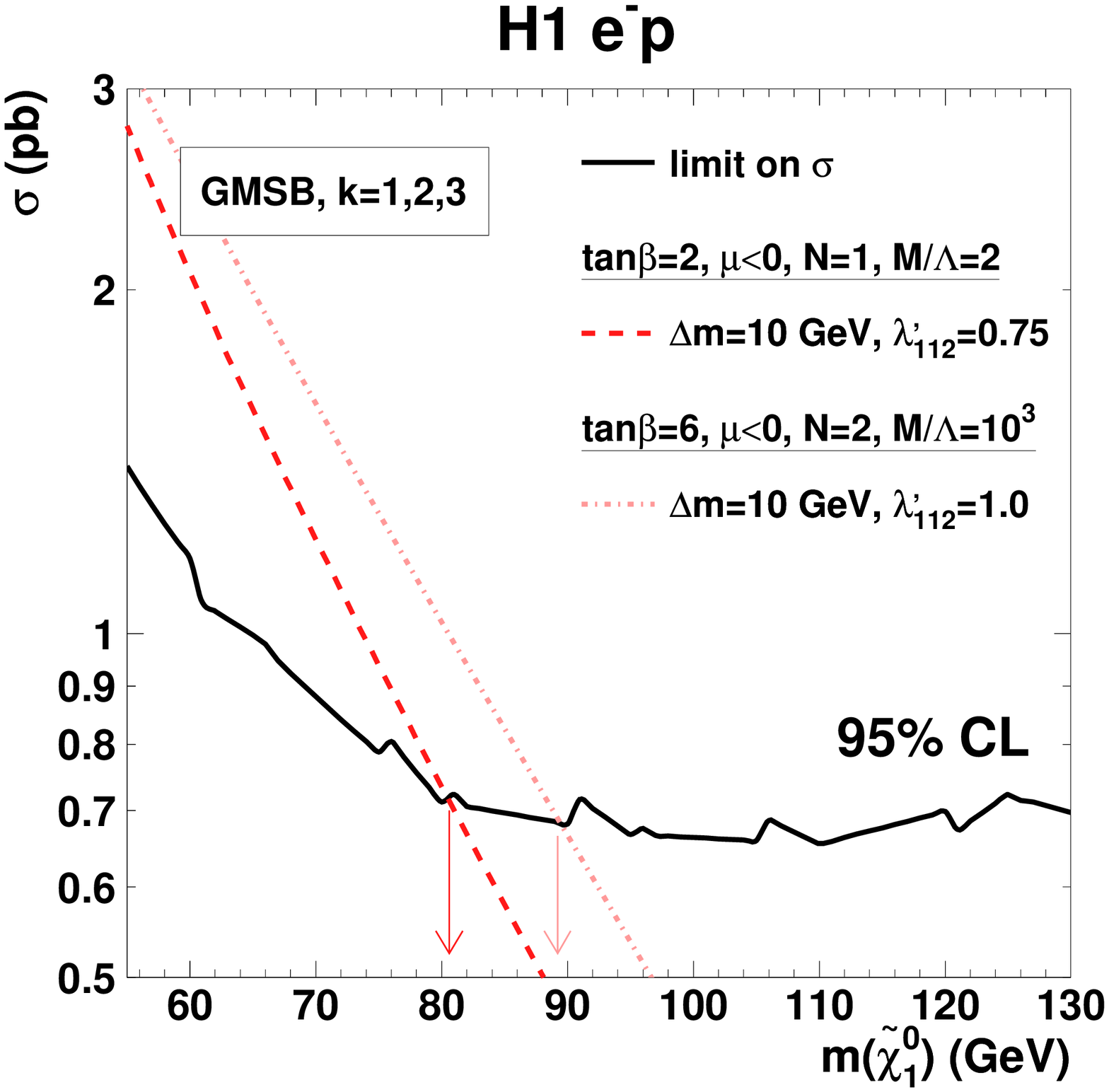,width=8cm,bbllx=25,bblly=660,bburx=575,bbury=130}}
\end{picture}
  \end{center}
  \vspace{8.5cm}
  \caption{Upper limit at the $95\,\%$ CL on the cross section as a function
    of the neutralino mass for example GMSB scenarios (solid lines). For
  comparison, the GMSB
    cross sections for different
    \rpv\ couplings $\lambda'_{121}$ and $\lambda'_{112}$
    are superimposed for a mass difference of $\Delta m =
  m({\tilde{e}_L})-m({\neu}) = 10 \,\GeV$ (dashed and dashed-dotted lines).}
  \label{fig:lambda}
\end{figure}

In Fig.~\ref{fig:lambda}, upper limits on
the cross sections are shown as a function of $m(\neu)$. 
The limits become less stringent at low
neutralino masses due to the reduced signal detection efficiencies.
The GMSB cross sections
for different values of the couplings $\lambda'_{121}$ and $\lambda'_{112}$
are also shown for a mass difference $\Delta m =
m({\tilde{e}_L})-m({\neu}) = 10\,\GeV$.
%%

%
%%%%%%%%%%%%%%%%%%begin%%figure%%%%%%%%%%%%%%%%%%%%%%%%%%%%%%%%%%%%%%%
%%%%%%%%%%%%%%%%%%%%%%%%%%%%%%end%%figure%%%%%%%%%%%%%%%%%%%%%%%%%%%%%
%
%%%%%%%%%%%%%%%%%%begin%%figure%%%%%%%%%%%%%%%%%%%%%%%%%%%%%%%%%%%%%%%
\begin{figure}[hhh] 
\vspace{-11cm}
  \begin{center}
\begin{picture}(0,100)
         \put(-82,-15){\epsfig{file=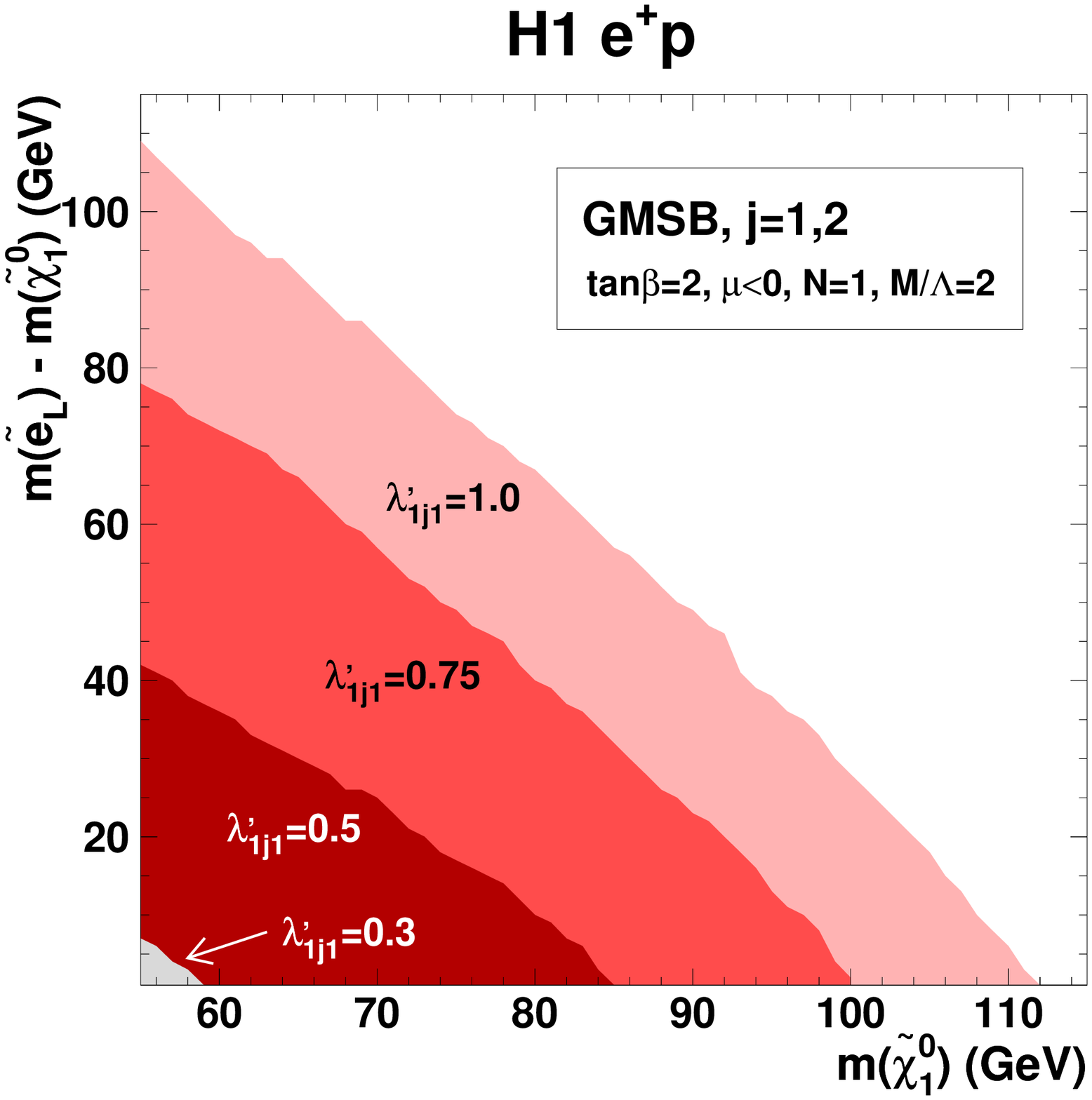,width=8cm,bbllx=25,bblly=660,bburx=575,bbury=130}}
         \put(3,-15){\epsfig{file=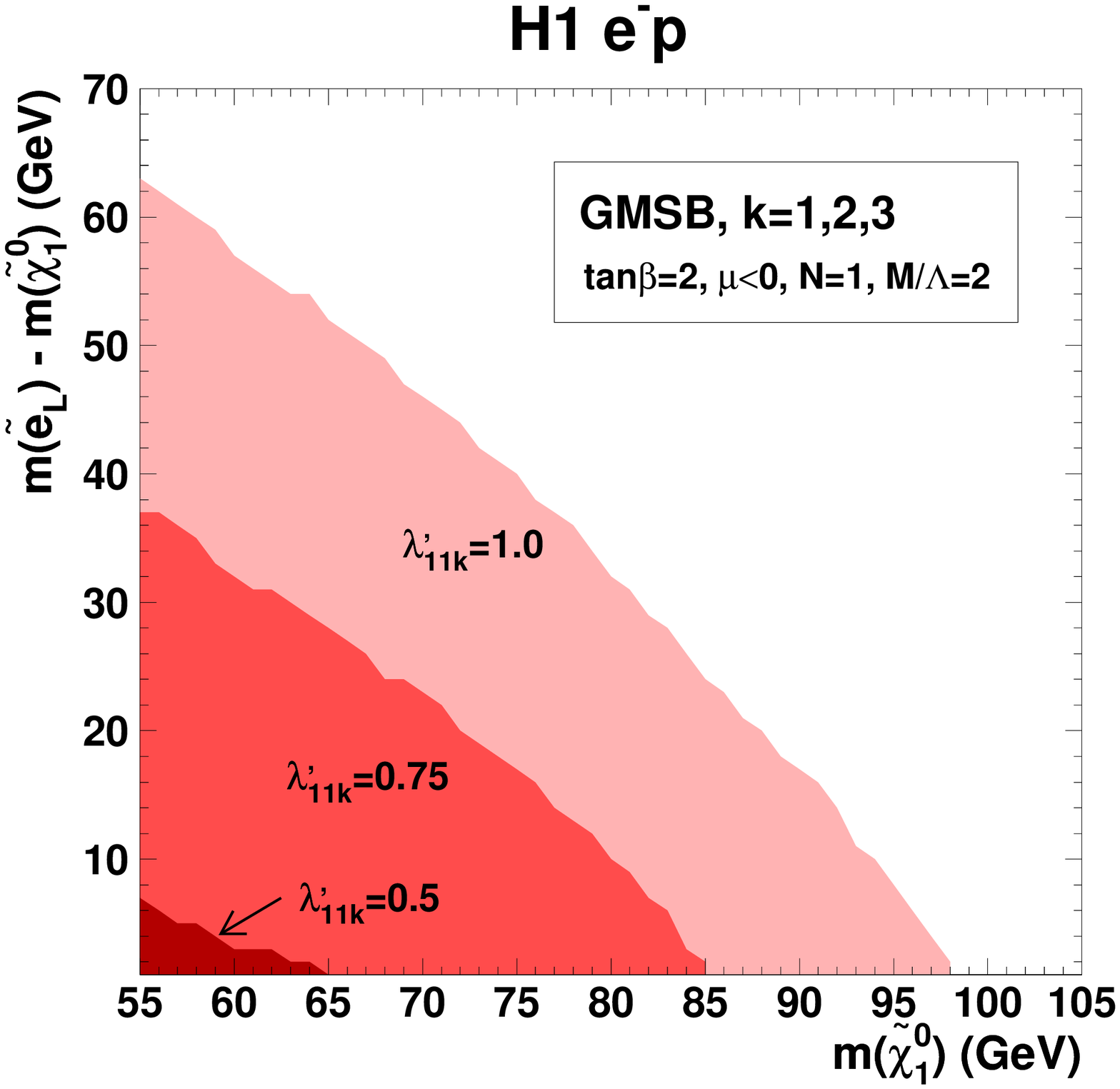,width=8cm,bbllx=25,bblly=660,bburx=575,bbury=130}}
         \put(-82,-100){\epsfig{file=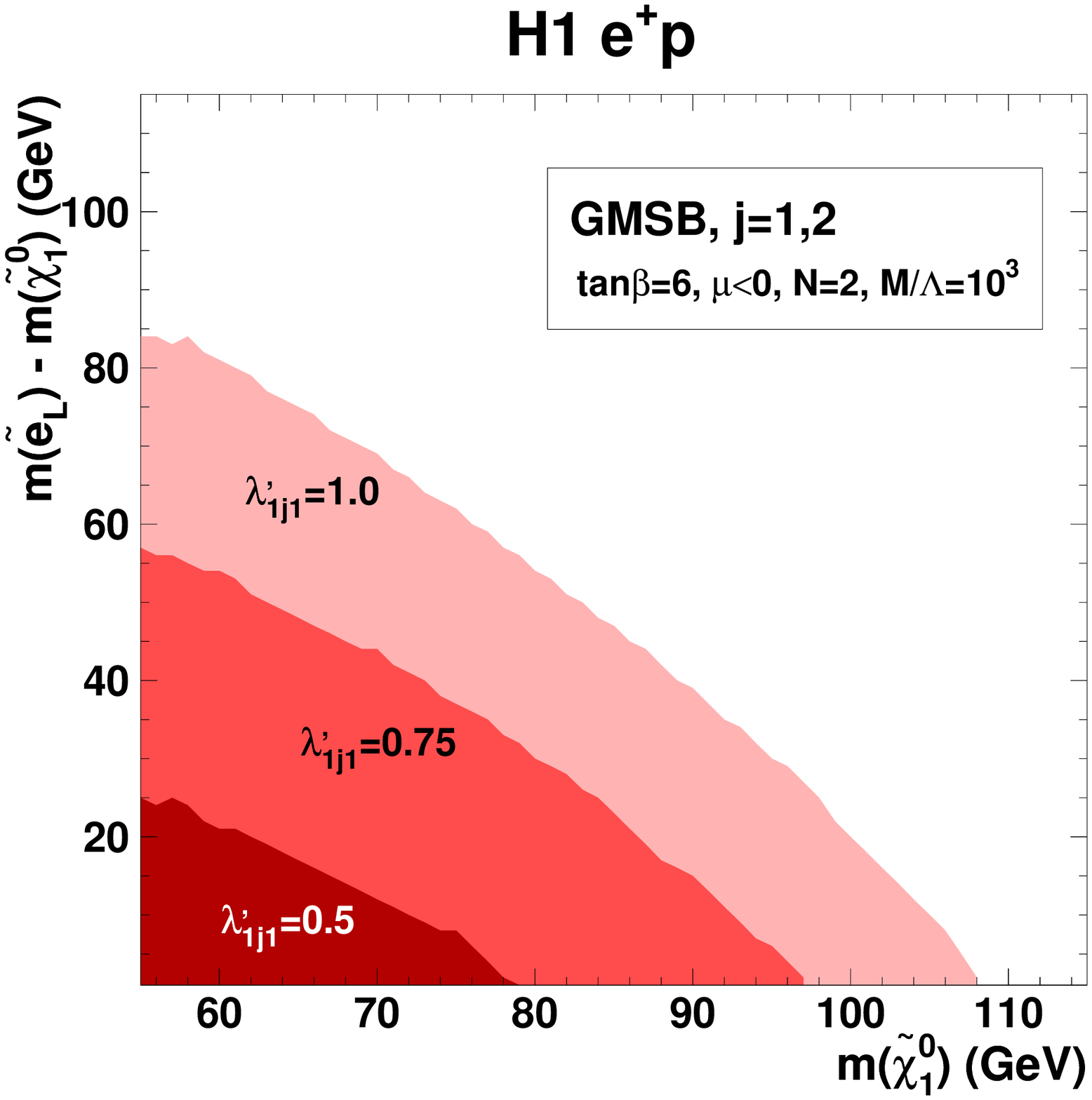,width=8cm,bbllx=25,bblly=660,bburx=575,bbury=130}}
         \put(3,-100){\epsfig{file=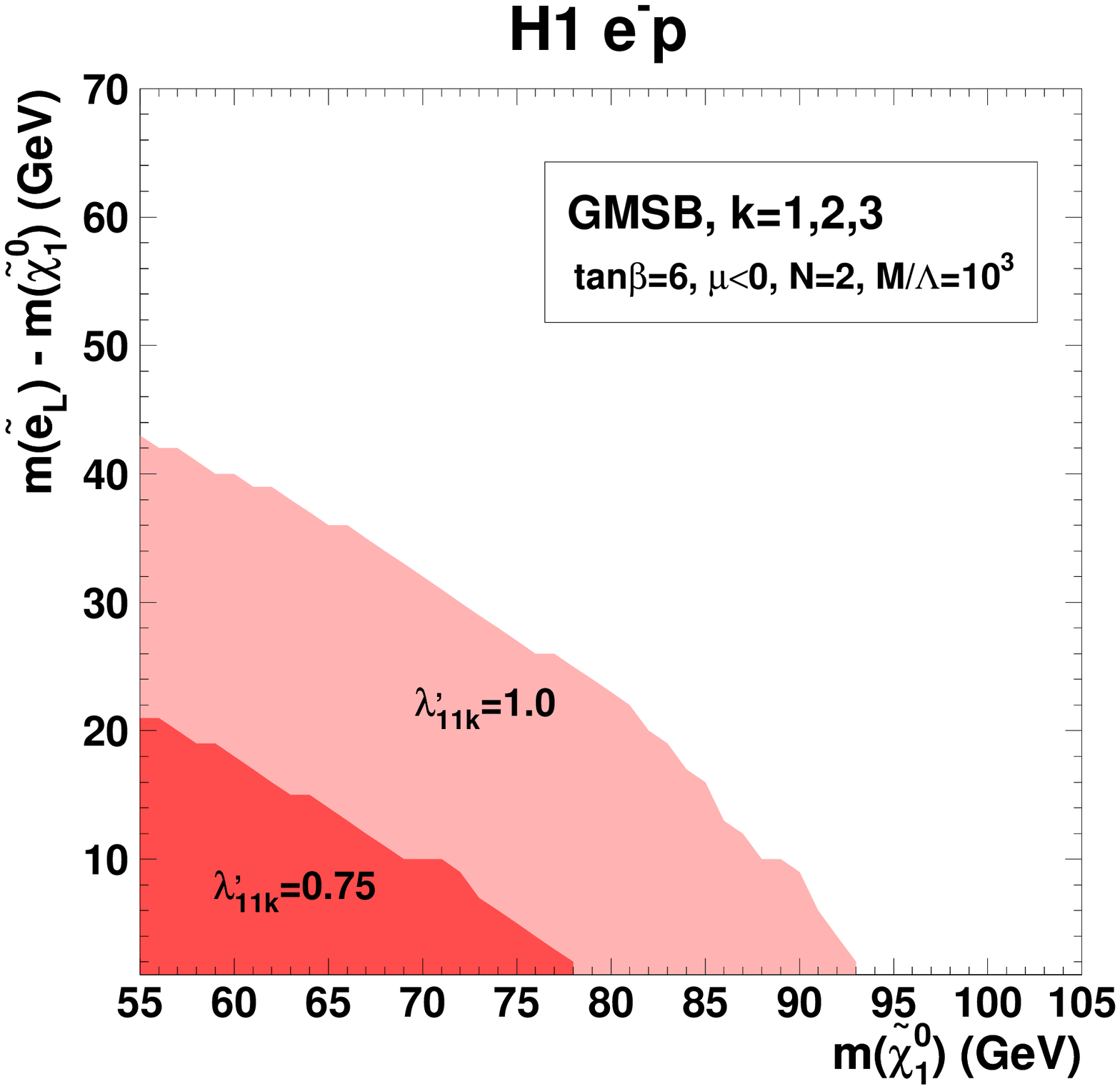,width=8cm,bbllx=25,bblly=660,bburx=575,bbury=130}}
\end{picture}
  \end{center}
  \vspace{17cm}
  \caption{Excluded regions at the $95\,\%$ CL in the
    $\Delta m = m({\tilde{e}_L})-m({\neu})$ and $m({\neu})$ plane
for various values of $\lambda'_{1j1}$ ($j=1,2$) and $\lambda'_{11k}$
($k=1,2,3$).
}
  \label{fig:mselmne}
\end{figure} 
%%%%%%%%%%%%%%%%%%%%%%%%%%%%%%end%%figure%%%%%%%%%%%%%%%%%%%%%%%%%%%%%

%
In Fig.~\ref{fig:mselmne}, excluded regions are  
presented in the plane spanned by
$\Delta m $ and $m({\neu})$
using data from  $e^+p$
and $e^-p$ collisions for various values for the respective \rpv\ coupling.
The excluded
domains, obtained for $\lambda'_{1j1}$ ($j=2$) and $\lambda'_{11k}$ ($k=2,3$),
conservatively apply also in the case of a $\lambda'_{111}$ coupling.
For $\tan\beta=2$, $N=1$ and $\lambda'_{1j1} = 1.0$, the $e^+p$ results
exclude neutralino masses up to
$112\,\GeV$ for small $\Delta m$. For
large $\Delta m$ and small neutralino masses, selectron masses up to $164\,\GeV$
are excluded. In $e^-p$ collisions, for $\tan\beta=2$, $N=1$ and 
$\lambda'_{11k} = 1.0$, neutralino masses up to
$98\,\GeV$ for small $\Delta m$ and selectron masses up to $118\,\GeV$ for
large $\Delta m$ are ruled out.

Apart from the coupling $\lambda'_{111}$, which is tightly constrained
by searches for the neutrinoless double beta decay of nuclei \cite{nldbd},
values of $\lambda'_{1jk} = 1$  ($(j,k) \ne (1,1)$) are not excluded
by indirect measurements when the squark masses are very
high\footnote{
      For example \cite{apv}, searches for atomic parity violation allow
      couplings $\lambda'_{1jk} = 1$ if the supersymmetric partner
      of the left-handed up-type (right-handed down-type) quarks
      are heavier than $3.5\,\TeV$ ($5\,\TeV$).
     }.
The limits on the $\lambda'_{121}$, $\lambda'_{112}$ and $\lambda'_{113}$
couplings obtained in this analysis are the first constraints which
depend only on the slepton and neutralino masses.
For instance, 
for masses of the $\neu$ and $\sell$ close to
$55\,\GeV$, couplings $\lambda'_{1j1}>0.3$ ($j=1,2$) and $\lambda'_{11k}>0.5$
($k=1,2,3$) are ruled out for
$\tan\beta=2$ and $N=1$. 

The range of neutralino masses which is excluded by this analysis
for \rpv\ couplings of the order of one is comparable with that which
is probed at the Tevatron \cite{Abazov:2004jx} and at LEP \cite{lepgmsb}.
It should be stressed, however, that our results are complementary
to those derived at the Tevatron where the dominant contributions
to the cross section are from the production of the lightest charginos
(${\tilde{\chi}_1^+}
{\tilde{\chi}_1^-}$) and chargino-second neutralino pairs
(${\tilde{\chi}_2^0} {\tilde{\chi}_1^\pm}$).
They are also
complementary to
those obtained at LEP since the process $e^+ e^- \rightarrow
\neu \neu \rightarrow \gamma \grav \gamma \grav$ is
mainly sensitive to the supersymmetric partner of the right-handed
electron.

%%%%%%%%%%%%%%%%%%%%%%%%%%%%%%%%%%%%%%%%%%%%%%%%%%%%%%%%%%%%%%%%%%%%%%%%
%%%%%%%%%%%%%%%%%%%%%%%%%%%%%%%%%%%%%%%%%%%%%%%%%%%%%%%%%%%%%%%%%%%%%%%%
\section{Conclusions}
%%%%%%%%%%%%%%%%%%%%%%%%%%%%%%%%%%%%%%%%%%%%%%%%%%%%%%%%%%%%%%%%%%%%%%%%
%%%%%%%%%%%%%%%%%%%%%%%%%%%%%%%%%%%%%%%%%%%%%%%%%%%%%%%%%%%%%%%%%%%%%%%%

Events containing a photon, a jet and large missing
transverse momentum are analysed in data from $e^\pm p$ collisions at
a centre-of-mass energy of 
$\sqrt{s}=319\,\GeV$ using the H1 
detector at HERA. Within the SM this topology is mainly produced by charged
current processes with photon radiation. Such events also arise in
gauge mediated SUSY breaking models with
$R$-parity violation (\rpv).
The data analysis reveals no deviation from the SM.
Constraints on GMSB models are derived for different values
of the \rpv\ coupling. 
For small mass differences between the neutralino $\neu$ and 
the supersymmetric partner of the left-handed electron $\sell$,
neutralinos with $m(\neu)$ up to $112\,\GeV$ are ruled out at the
$95\,\%$ CL for \rpv\
couplings $\lambda'=1$. 
Similarly, for large mass differences,
masses $m(\sell)$
up to $164\,\GeV$ are excluded.
For masses $m(\neu)$ and $m(\sell)$ close
to $55\,\GeV$, $\lambda'_{1j1}$ Yukawa couplings of electromagnetic
strength are excluded. These are the first constraints from HERA on SUSY
models which are independent of the squark sector.

%%%%%%%%%%%%%%%%%%%%%%%%%%%%%%%%%%%%%%%%%%%%%%%%%%%%%%%%%%%%%%%%%%%%%%%%
%%%%%%%%%%%%%%%%%%%%%%%%%%%%%%%%%%%%%%%%%%%%%%%%%%%%%%%%%%%%%%%%%%%%%%%%
%
\section*{Acknowledgements}

We are grateful to the HERA machine group whose outstanding
efforts have made this experiment possible. 
We thank
the engineers and technicians for their work in constructing and
maintaining the H1 detector, our funding agencies for 
financial support, the
DESY technical staff for continual assistance
and the DESY directorate for support and for the
hospitality which they extend to the non-DESY 
members of the collaboration. We would like to thank V.~B{\"u}scher,
K.~P.~Diener, C.~Rembser and M.~Spira for valuable discussions.

%%%%%%%%%%%%%%%%%%%%%%%%%%%%%%%%%%%%%%%%%%%%%%%%%%%%%%%%%%%%%%%%%%%%%%%%
%%%%%%%%%%%%%%%%%%%%%%%%%%%%%%%%%%%%%%%%%%%%%%%%%%%%%%%%%%%%%%%%%%%%%%%%
%
%   References for Gravitino paper
%

%
%
%%%%%%%%%%%%%%%%%%%%%%%%%%%%%%%%%%%%%%%%%%%%%%%%%%%%%%%%%%%%%%%%%%%%%%%%
%%%%%%%%%%%%%%%%%%%%%%%%%%%%%%%%%%%%%%%%%%%%%%%%%%%%%%%%%%%%%%%%%%%%%%%%

\end{document}